\documentclass[aps,twocolumn,showpacs,prl,amsmath,amssymb,floatfix,superscriptaddress]{revtex4-1}
\usepackage[utf8]{inputenc}
\usepackage{amsmath}
\usepackage{amssymb}
\usepackage{braket}
\usepackage{comment}
\usepackage{bbold}
\usepackage{graphicx,color}
\DeclareMathOperator{\Tr}{Tr}

\def\be{\begin{equation}}
\def\ee#1{\label{#1}\end{equation}}

\def\b{\beta}
\def\g{\gamma}

\def\d{\delta}

\def\l{\lambda}

\def\m{\mu}

\def\x{\xi}
\def\p{\pi}
\def\r{\rho}

\def\f{\varphi}

\def\o{\omega}

\def\i{\int}

\def\bv{{\mathbf v}}

\def\bz{{\mathbf z}}
\def\bZ{{\mathbf Z}}

\def\bxi{\bm\x}

\begin{document}

\title{Comment on ``Measurability of nonequilibrium thermodynamics in terms of the Hamiltonian of mean force''}
\author{Peter Talkner}
\affiliation{Department of Physics, University of Augsburg, D-86135 }  
\author{Peter H\"anggi}
\affiliation{Department of Physics, University of Augsburg, D-86135 Augsburg, Germany}
\affiliation{ Nanosystems Initiative Munich, D-80799 M\"unchen, Germany}
\affiliation{Department of Physics, National University of Singapore, Singapore 117546}
\date{\today}

\begin{abstract} 
In a recent paper [P. Strasberg and M. Esposito, Phys. Rev. E {\bf 101}, 050101(R) (2020)] 
an attempt is presented to formulate the nonequilibrium thermodynamics of an open system in terms of the Hamiltonian of mean force. The purpose of the present comment is to clarify severe restrictions of this approach and also to stress that recently noted ambiguities [P. Talkner and P. H\"anggi, Phys. Rev. E {\bf 94}, 022143 (2016)] of fluctuating thermodynamic potentials cannot be removed in the suggested way.              
\end{abstract}

\maketitle
The Hamiltonian of mean force~\cite{CTH09}, like its classical precursor, the potential of mean force~\cite{Onsager,Kirkwood}, encompasses the complete information on the equilibrium thermodynamical behavior of an open system. It is defined in terms of the bath-renormalized Boltzmann factor at the inverse temperature $\b$, given by the expression 
\begin{equation}
e^{-\beta H^*(\b,\m)} = Z^{-1}_B(\b,\m) \Tr_B e^{-\beta H_{\text{tot}}(\m)}\:,
\label{Hmf}
\end{equation}
where $H^*(\b,\m)$ and $H_{\text{tot}}(\m) = H_S(\m)+H_{SB}(\m) +H_B(\m) $ denote the Hamiltonian of mean force and Hamiltonian of the system, $H_S(\m)$, and the environment, $H_B(\m)$, including the mutual interaction, $H_{SB}(\m)$, respectively. With $\m$, all externally controllable parameters, including those parameters $\l$, exclusively acting on the considered open system, but also including all other globally acting parameters, such as electric and magnetic fields, are specified. Furthermore, $\Tr_B$ indicates the partial trace over the environmental Hilbert space for quantum systems and, for a classical system, the integral over the environmental phase space with respect to a 
properly permutation-symmetry-adapted dimensionless volume element. 
With the normalization by the partition function of the environment, $Z_B(\b,\mu) = \Tr_B e^{-\b H_B(\m)}$, the Hamiltonian of mean force agrees with the bare system Hamiltonian for isolated systems;  for further details see~\cite{TH16,TH19}. With the partition function for the Hamiltonian of mean force $Z_S(\b,\m) = \Tr_S e^{-\b H^*(\b,\m)} = Z_{\text{tot}}(\b,\m)/Z_B(\b,\m)$ and  the resulting Helmholtz free energy $F_S(\b,\m) = -\b^{-1} \ln Z_S(\b,\m)$ the equilibrium thermodynamics of the open system becomes accessible. Likewise, for a quantum system, the density matrix and, for a classical system, the respective phase space probability density function (pdf)  are specified in terms of the Hamiltonian of mean force according to
\be
\r(\b,\m) = Z^{-1}_S(\b,\m) e^{-\b H^*(\b,\m)}\:.
\ee{rb}
On the other hand, from the knowledge of $\r(\b,\m)$, which can in principle be inferred from a system-intrinsic point of view, it is not possible to separate $\r(\b,\m)$ into the Hamiltonian of mean force and the system's partition function in conflict with a central dogma of stochastic thermodynamics~\cite{Seifert12,Seifert19,SE} which postulates that all necessary information can be obtained from observations of the system without recourse to data from the environment. As a resort, Strasberg and Esposito argue in~\cite{SE} that only those differences of free energies, or equivalently, those  ratios of system's partition functions are relevant for the description of thermodynamic processes that are taken at different values of the direct system control parameters $\l$. 
With this assumption, 
the large class of thermodynamic processes accompanied by  changes of  global parameters as well as  of temperature and pressure is excluded and most relevant thermodynamic quantities such as specific heat, magnetization and polarization, magnetic and electric susceptibilities and compression factors are not accessible within such an approach.  Hence, even though the description of special processes  
may not require extra information about the environment, 
in general, a system intrinsic description, i.e., one that is exclusively based on the observation of the system, be it by means of quantum tomography or of a monitoring of the stochastic trajectories of a classical open system, is not possible~\cite{TH16,TH19}. In contrast to Strasberg and Esposito~\cite{SE}, being concerned with the nonequilibrium dynamics of a special class of problems, our main aim in \cite{TH16,TH19} is the characterization of the thermal equilibrium of open systems  in its dependence on {\it all} relevant parameters. In this context, the modification of the Hamiltonian of mean force suggested by Strasberg and Esposito cannot be qualified as an addition of an ``irrelevant constant'' that that would have ``{\it no} thermodynamic consequences''; the quotations are taken from~\cite{SE}. In general it would rather lead to erroneous conclusions.
Even in the restricted situation of isothermal processes at constant global parameters it is not sufficient to know the reduced density matrix or the pdf of the open system to infer changes of thermodynamic quantities. Additionally, process-specific relations such as the Jarzynski equality  or second-law-like relations must be imposed. This  further narrows the predictive power of the method suggested by~\cite{SE}. Moreover,  the inference of free energy differences on the basis of the Jarzynski equality often requires an unrealistically large amount of data~\cite{Kim,Deng}. The same kind of problems must be expected also for the other methods, in particular, when the  tomography of time-dependent states is required to estimate e.g. an effective system Hamiltonian on the basis of Eq. (23) in~\cite{SE}.

As explained in~\cite{TH16,TH19}, the concept of fluctuating thermodynamic potentials suffers already in thermal equilibrium from ambiguities that can be subsumed as the set of all functions with a vanishing equilibrium average value. For transient and other nonequilibrium processes, the respective set of functions is characterized by a vanishing average with respect to the actual, time-dependent state of the system. It hence changes with time 
but it does not collapse to an empty set as Strasberg and Esposito wrongly conclude in the footnote [60] of reference ~\cite{SE}. We further note that even the specification of a particular fluctuating free energy, as done in Eq. (5) of~\cite{suppSE}, does still leave the fluctuating internal energy and entropy largely unspecified, as can be seen from the Eqs. (107), (108) in~\cite{TH19}. In the particular case of an equilibrium system the supposedly fluctuating free energy defined in (5) of~\cite{suppSE} yields the non-fluctuating equilibrium free energy~\cite{TH16}.     

Furthermore, we would like to clarify two misleading literal citations in~\cite{SE} which are taken out of their original context. Our statement in \cite{TH19} that ``...presents in practice an impossible task'' does not refer to the Hamiltonian of mean force as insinuated by Strasberg and Esposito~\cite{SE} but to the reconstruction of the total system's Hamiltonian solely based on open system's trajectories. Our observation that the first law of thermodynamics for quantum open systems interacting with their environments at a finite strength is doubtful is based on the fact that then the respective observables that  determine work and heat do not commute. Hence, their simultaneous measurement is excluded by the laws of quantum mechanics. The  condition that measurements ``need to be error free'', cited from \cite{TH19},  
refers to one of the mathematical properties that  generalized energy measurements must satisfy in order to yield the Crooks and the Jarzynski fluctuation relations~\cite{VWT,ITVW}, but has not been made in the context of the first law.        

Finally, we would like to stress that, by its very definition, the Hamiltonian of mean force 
describes thermodynamic {\it equilibrium}. Expressions like the nonequilibrium free energy in Eq. (5) of~\cite{SE} or the corresponding fluctuating nonequilibrium free energy in Eq. (5) of~\cite{suppSE}, based on the Hamiltonian of mean force and the pdf at the time $t$ and at the position of a random trajectory,  
are mere postulates without a deeper rational. 
The latter object, resulting in the described way from the pdf,  is by construction a function of the starting point of the considered random trajectory and a functional of the random force having acted up to the time $t$. This very construct 
does not follow as a  transformation of a pdf according to the proper rules of probability theory. It therefore
is not normalized with respect to the starting point and, hence, has no obvious probabilistic meaning~\cite{supp}.




\onecolumngrid
\section{Supplementary Material}
\subsection{The ``fluctuating probability density function'' of stochastic thermodynamics}
The theory of stochastic thermodynamics is based on the hermaphroditic notion of a fluctuating probability density function (fpdf) whose functional form results from a {\it proper} probability density function (pdf) in which the state space variable is substituted by a particular realization of the considered  process~\cite{S06,S12,SE}. In the present supplemental material we would like to illustrate this construct, combining aspects of states and observables as well as forward and backward dynamics,  with the example of Markovian diffusion processes.
Finally we shall specialize to the case of a Brownian oscillator.
If we denote a point in the state space of the considered process as $\bz$, the pdf, which is a solution of the Fokker-Planck equation, by $\r(\bz,t)$ and a solution of the Langevin equation by $\bZ(\bz,t)$ with $\bZ(\bz,0)=\bz$, the fpdf $\f(\bz,t)$ is defined according to~\cite{S06,S12,SE} as 
\be
\f(\bz,t) = \r(\bZ(\bz,t),t)\:,
\ee{fr} 
which is a function of the starting point of the considered realization and no longer of its end point. Trivially, at the initial time $t=0$ the fpdf and pdf agree with each other. In the sequel we demonstrate by means of the example of a Brownian oscillator that the fpdf does {\it not} stay normalized, in general, as its very construction does not conform with the transformation rules of probability densities.  An exception presents a Hamiltonian dynamics, as shown below.     
\subsection{Brownian harmonic oscillator}
\subsubsection{Langevin equation and the probability density function}   
The process of a damped harmonic oscillator of mass $m$ and frequency $\o$ under the influence of a Gaussian white random force $\x(t)$ is described by the Langevin equation \cite{WU}
\be
\begin{split}
\dot{q}(t) &= \frac{p(t)}{m} \\
\dot{p}(t) &= - \g p(t) - m \o^2 q(t) +\x(t)\:,
\end{split}
\ee{LE}
where $q(t)$ and $p(t)$ are the position and momentum, respectively, of the oscillator at the time $t$, and $\g$ denotes the friction constant. The average of the random force $\x(t)$ vanishes and its auto-correlation function is given by 
\be
\langle \x(t)\x(s) \rangle = 2 m \g k_B T \d (t-s),
\ee{db}
depending on the temperature $T$ of the bath, causing the frictional and fluctuating forces. Here, $k_B$ denotes the Boltzmann constant.  The vector $\bZ(\bz,t) = \big (Q(\bz,t), P(\bz,t) \big ) $ of the solutions  $Q(\bz,t)$ and $P(\bz,t)$ of Eq.~\ref{LE} starting at $\bZ(\bz,0) = \bz =(q,p)$ can be written as
\be
\bZ(\bz,t) = \bZ_h(\bz,t) + \bZ_i(t)\:,
\ee{Z}
where 
\be
\begin{split}
\bZ_h(\bz,t) &= e^{R t} \bz \\
\bZ_i(t) & = \i_0^t ds e^{R(t-s)} \bxi(s)\:.
\end{split}
\ee{ZhZi}
Here, $R$ is the matrix of the coefficients of $q(t)$ and $p(t)$ on the right hand side of Eq.~\ref{LE}, hence reading
\be
R = \left (
\begin{array}{cc}
0& 1/m\\
-m \o^2 & -\g
\end{array}
\right )\:.
\ee{R}
Accordingly, the exponentiated  matrix becomes
\be
e^{R t} = e^{-\g t/2} \left (
\begin{array}{cc}
\cos \o_\g t + \frac{\g}{2\o_\g} \sin \o_\g t &\frac{1}{m\o_\g} \sin \o_\g t\\
- \frac{m \o^2}{\o_\g} \sin \o_\g t & \cos \o_\g t  - \frac{\g}{2 \o_\g} \sin \o_\g t\:
\end{array}
\right )\:,
\ee{eRt}
where $\o_\g =(\o^2 - \g^2/4)^{1/2}$ is the effective frequency of the oscillator. For the sake of simplicity we assume $\o >\g/2$.   
Finally, $\bxi(t) = (0, \x(t))$ denotes the vectorial random force.

In order to construct the fpdf we need the time-dependent pdf $\r(\bz,t) $. 
Due to the linearity of the process defined by the Langevin equation (\ref{LE}), an initially Gaussian pdf stays Gaussian for all later times $t$, taking the form
\be
\r(q,p,t)= \left ( 2 \p M(t) \right )^{-1/2} e^{-\frac{1}{2} (\bz -\langle \bZ(t) \rangle )^{tr} \cdot M^{-1}(t) \cdot (\bz -\langle \bZ(t) \rangle )^T }
\ee{gauss}
with the superscript $tr$ indicating the transposed respective vector or matrix. According to the Eqs. (\ref{Z}) and (\ref{ZhZi}), the average $\langle \bZ(t) \rangle$ becomes
\be
\langle \bZ(t) \rangle = e^{R t} \langle \bz \rangle\:,
\ee{Zt}
where $\langle \bz \rangle$ denotes the average of $\bz$ with respect to the initial distribution. The time-dependence of the covariance matrix $M(t)$ of the vector $\bZ$ follows from the following equation of motion~\cite{Ludwig}
\be
\dot{M}(t) = M(t) R^{tr} + R M(t) +2 D
\ee{dM}
with $D$ denoting the diffusion matrix reading
\be
D = \left (
\begin{array}{cc}
0 & 0\\
0 & m \g k_B T
\end{array}
\right )\:,
\ee{D}
yielding for the time-dependent autocorrelation matrix the expression
\be
M(t) = e^{R t} \left [M(0)  +2 \i_0^t ds e^{-R s} D   e^{-R^{tr} s} \right ] e^{R^{tr} t}\:.
\ee{MtR}
\subsection{The ``Fluctuating probability density function''}
Combining the Eqs. (\ref{fr}) and (\ref{gauss}) with (\ref{Z},\ref{ZhZi}) and (\ref{MtR}) one obtains 
\be  
\f(\bz,t) = \left (2 \p M(t) \right )^{-1/2} e^{-1/2 (\d z - \bv(t))^{tr} Q^{-1}(t)(\d z - \bv(t)) }\:,
\ee{fho} 
where $\d \bz = \bz - \langle \bz \rangle_0$ denotes the fluctuations of the phase space points $\bz$ in the initial ensemble specified by $\r(\bz,0)$, $\bv(t) = \i_0^t ds e^{-R s} \bxi(s)$ and $Q(t) =M(0) + 2 \i_0^t ds e^{-R s} D e^{- R^{tr} s}$. 
The integral of the fpdf extended over all initial phase space points then results in
\be
\begin{split}
\i d\bz \f(\bz,t) &= \left [ \frac{\det Q(t) }{\det M(t)} \right ]^{1/2}\\
&= \det e^{-R t/2} = e^{\g t /2}\:,
\end{split}
\ee{ifpdf} 
where we used Eq. (\ref{eRt}). Therefore, the fluctuating probability density does in general not stay normalized to unity and hence cannot be interpreted as a probability density. We note that the time dependence of the integral in Eq. (\ref{ifpdf}) is due to the dissipation, but not due to the randomness of the Brownian oscillator dynamics. 

In the absence of friction, for purely Hamiltonian dynamics, the substitution of a time evolved trajectory in the probability density leads back to the initial pdf as can be seen by writing the time-evolved pdf $\r(\bz,t)$ in terms of the initial pdf $\r(\bz,0)$ according to the general transformation rules of pdfs as
\be
\r(\bz,t) =\i d\bz_0 \d(\bz-\bZ(\bz_0,t)) \r(\bz_0,t) =       \r(\bZ^{-1}(\bz,t), 0)\:,
\ee{rtr0}
with the Hamiltonian trajectory $\bZ(\bz,t)$, mapping the initial to the final phase space point and its uniquely defined inverse  $\bZ^{-1}(\bz,t)$ acting oppositely. Hence, the replacement of $\bz$ by the Hamiltonian trajectory $\bZ(\bz,t)$ in the time-dependent pdf leads back to the initial pdf according to
\be
\f(\bz,t) = \r(\bZ(\bz,t),t) = \r(\bz,0)\:.
\ee{frtr0}
       

\end{document}